\documentclass[journal=jacsat,manuscript=article]{achemso}

\usepackage{chemformula} 
\usepackage[T1]{fontenc} 
\usepackage{subfigure}    

\usepackage{ulem}
\usepackage{booktabs}

\author{Weihang Gao}
\author{Teng Zhao}
\affiliation{School of Mathematical Sciences, Shanghai Jiao Tong University, Shanghai 200240, China}
\author{Shian Dong}
\author{Xingyi Huang}
\affiliation{Department of Polymer Science and Engineering, Shanghai Key Laboratory of Electrical Insulation and Thermal Aging, State Key Laboratory of Metal Matrix Composites, Shanghai Jiao Tong University, Shanghai 200240, China}
\author{Zhenli Xu}
\email{xuzl@sjtu.edu.cn}
\affiliation{School of Mathematical Sciences, CMA-Shanghai, MOE-LSC and Shanghai Center for Applied Mathematics, Shanghai Jiao Tong University, Shanghai 200240, China}

\title{Microscopic Energy Storage Mechanism of Dielectric Polymer-Coated Supercapacitors}

\abbreviations{IR,NMR,UV}

\begin{document}

%
%
%

\begin{abstract}
Supercapacitors have been attracting significant attention as promising energy storage devices. However, the voltage window limitation associated with electrolyte solutions has hindered the improvement of their capacitance. To address this issue and enhance the energy storage capabilities of general traditional supercapacitors, we put forward the dipole-induced effects observed in the theoretical framework of the electric double-layer structure. The molecular dynamics results demonstrate that, compared to traditional systems, an improvement of over 50$\%$ in integral capacitance at low voltages is achieved. Moreover, a new material-based experimental results obtained from a dielectric supercapacitor employing a hydrated electrolyte solution corroborated the effectiveness of our proposed model, yielding consistent outcomes. We attribute the large capacitance variation to the reorientation of the dipoles, which induces the neutral-to-bilayer transition and the overscreening-to-steric transition, consistent with the polarization process of the polymer in the experiment.  We further investigate the capacitance variations under different dipole parameters, such as varying the number of layers, different number densities and different spacings, thereby enriching the experimental results with additional conclusions not previously obtained. This work presents a novel approach that exploits dipole-induced capacitance effects, paving the way for further advances in the field of energy storage technology.
 
\end{abstract}

\section{Introduction}
High-performance energy storage issue is becoming increasingly significant due to the accelerating global energy consumption \cite{liu2018advanced,noori2019towards,simon2020perspectives}. Among various energy storage devices\cite{winter2004batteries,simon2014batteries}, supercapacitors have attracted considerable attention owing to many outstanding features such as fast charging and discharging rates, long cycle life, and high power density\cite{zhao2023molecular,fan2023high,bhat2023frontiers,satpathy2023depth}. Compared with supercapacitors with ionic liquid  \cite{yu2019ionic}\cite{wang2023unraveling} or organic electrolytes\cite{sharma2019review}\cite{yeletsky2022today}, aqueous supercapacitor has been considered as a promising choice attributed by its low cost, high security, environmental friendliness, high accessibility and many other  advantages\cite{yu2017boosting}\cite{guo2022perspectives,gajewska2023effect}. However, the maximum operating potential ($\sim 1.23V$) for decomposing water\cite{jabeen2017high} becomes a bottleneck challenge for aqueous supercapacitors to achieve higher energy density (or the capacitance)\cite{xiong2018harmonizing}. One of the crucial ideas to tackle this problem is the utilization of the `water-in-salt' electrolyte, in which the highly concentrated ionic liquids prevent water from splitting at high operating voltages \cite{suo2015water,mceldrew2018theory}. Designing different electrode materials for asymmetric supercapacitors is another important strategy to enhance their energy density by promoting redox reactions that take precedence over water decomposition.\cite{xiong2018harmonizing,sahoo2018redox}. Beyond that, through the interface engineering\cite{zhao2021atomic} including modifications like polymer coating\cite{dong2020electrodes}\cite{chen2022conductive,shen2022facile}and surface roughness \cite{jia2018heterostructural,slesinski2018self} on the electrodes, the energy storage performance can also be enhanced.


Theoretical investigations into the factors that influence capacitance of supercapacitors have been well documented. Kornyshev\cite{kornyshev2007double} and Bazant {\it et al.} \cite{kilic2007steric,bazant2009towards} made pioneering contribution by using the lattice-gas model incorporated to the modified Poisson-Boltzmann equation to investigate differential capacitance for the case of symmetric 1:1 electrolytes near a planar electrode. They have uncovered that the volume occupancy effect plays a crucial role in influencing the variation of capacitance. Furthermore, the electrochemical performance is affacted by many other factors, such as ion adsorption \cite{uematsu2018effects}, ion hydration \cite{caetano2016role}, and dielectric decrement \cite{nakayama2015differential,qing2020effects}.  Especially, the dielectric change arising from the reorientation of dipoles in water\cite{schlaich2016water} has aroused great interest. Several studies have reported that the reorientation of dipoles can significantly affect the electric double layer (EDL) structure and capacitance\cite{mondal2020water,jiang2012solvent,zhan2017computational}. For example, Jiang and Wu\cite{jiang2013microscopic} reported that the integral capacitance with dipolar organic solvent is larger than that in ionic liquids. However, the dipolar effect near the charged electrode  on the capacitance remains less understood.

Considering the spatial effects and dipole effects mentioned above, our research aims to explore the impact of dipoles in the vicinity of the electrodes of supercapacitors on the capacitance. In this paper, the dipoles are positioned close to the electrode surface and allowed to rotate exclusively around their center of mass to consider the effect of reorientation.

\section{Methods}



To accurately model the physical mechanisms of dipole-induced effects for different solution systems and to simplify the simulation experiments, we employ a primitive model, in which the solvent is the relative dielectric constant\cite{lamperski2015planar} by molecular dynamics (MD) simulation. Specifically, we utilize the relative dielectric permittivity $\varepsilon_r=44.4$ to represent the water solvent. This choice of dielectric constant has been demonstrated to be appropriate for modeling high concentrations of potassium chloride electrolyte\cite{gavish2016dependence}. In the primitive model utilized in our study, the solvation effect can be accounted for by adjusting the ion sizes in the primitive model. Additionally, the resistance from the solvent and interface transfer\cite{merrill2009electrolyte,west2016reduction} have a negligible influence on the charge accumulation within the dipole layer near the electrode plates. Therefore, in this work, we have disregarded the solvent resistance and interface transfer resistance.

The dimensionless variables are used throughout the paper.
The length unit takes $d=0.36 nm$ (the diameter of the chloride ion \cite{nightingale1959phenomenological}). We take the ion mass $m=1$ and the energy unit $\varepsilon_{LJ}=k_BT$ which means the reduced temperature $T^{*}=1$. Here $k_B$ is the Boltzmann constant, the absolute temperature takes $T=300K$, and $\beta=1/k_BT$. The dimensionless physical quantities $z^{*}=z/d, \rho^{*}=\rho d^3, Q^{*}=Qd^2/e,  \varPhi^{*}=\beta e\varPhi, \sigma^{*}_s=\sigma_sd^2/e,  C^{*}=C\beta d/e^2$ and $v=\Omega_{ion}/\Omega$ represent the reduced position,  density,  charge, electrical potential, surface charge density, capacitance and the reduced ion volume ratio, respectively. Here, $e$ is the elementary charge, $\Omega_{ion}$ is the volume occupied by the ions and $\Omega$ is the volume of the simulation system. The unit of time $t^{*} = \sqrt{d^2m/\varepsilon_{LJ}}=1$. In the following, the asterisks have been omitted for the sake of simplicity.

\begin{figure}
	\hspace{-1.0em}
	\includegraphics[width=1.0\linewidth]{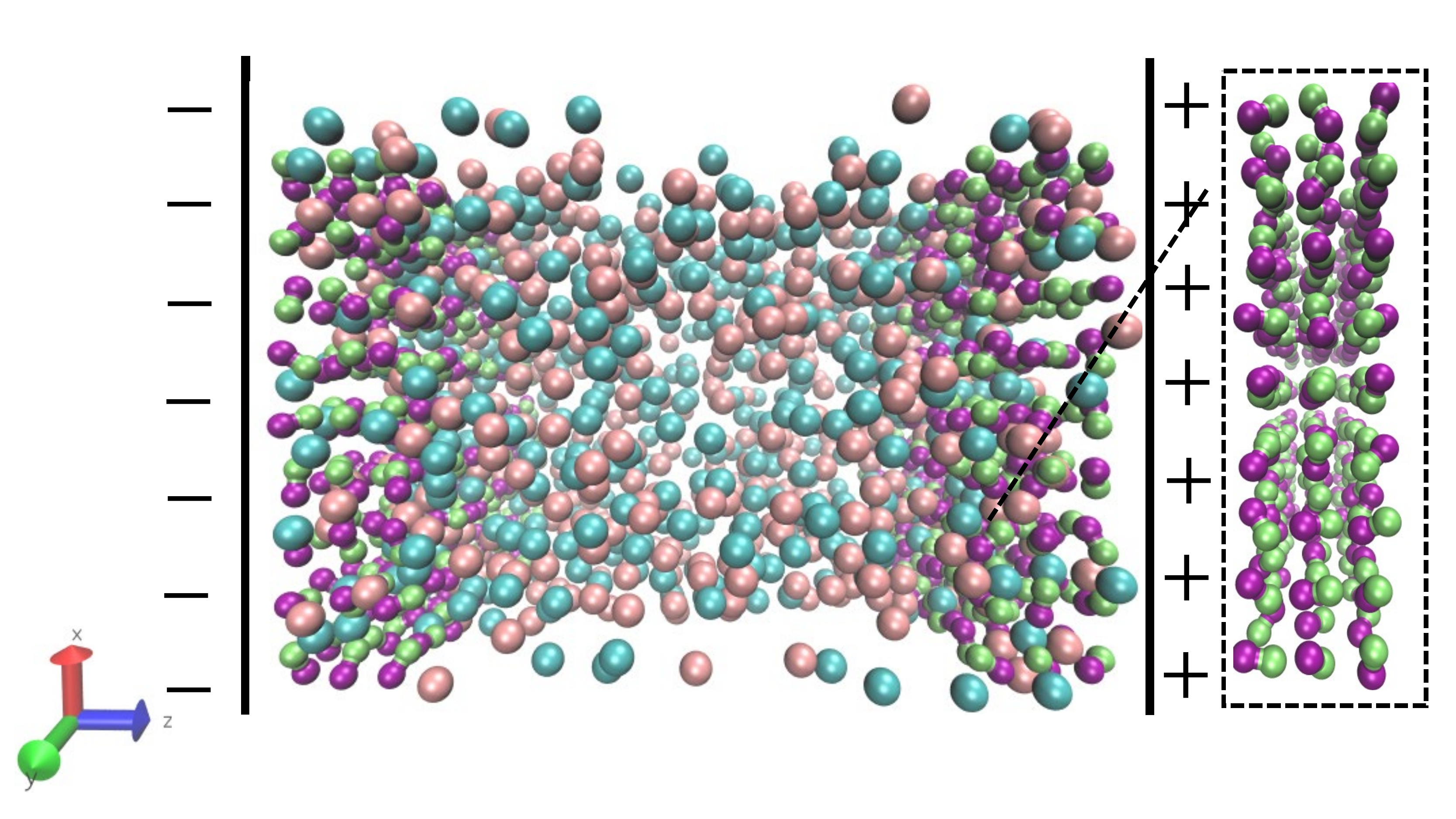}
	\caption{Schematic diagram of the simulation system with anions(pink spheres), cations(cyan spheres) and three layers of fixed dipoles(purple spheres connecting the lime spheres) shown on the right. The voltages applied to the electrodes are represented by the signs ``+'' and ``-'' respectively. }
	\label{molecular_system}
\end{figure}

The dipole is modeled as two hard spheres of opposite charges connected with a rigid rod. The center of the dipole is fixed while both ends are free to rotate. Periodic boundary conditions are applied to the $x$ and $y$ axes. The dimensions of the simulation box are set as $L_x = L_y = 20$ and $L_z = 30$. The electrodes are positioned at $z = \pm L_z/2$, respectively. Each electrode carries a fixed surface charge $Q_s$, resulting in a constant force of $F = 4\pi Q_s q / (L_x L_y \varepsilon_c)$ acting on each ion $q$. The diameter of each sphere in a dipole is set to 0.6, and the rod length to 0.3, with each layer containing the same number of dipoles. In the case of a three-layer system, the dipole layers are positioned at $z = -13.95$, $-12.15$, and $-10.35$ along the $z$ axis, respectively. To ensure the dipoles remain nontranslational, the net force on their translational degrees of freedom is maintained at zero. Additionally, the "Shake" option is employed to constrain the bond length of each dipole. The time step $\Delta t$ for the MD simulation is set to 0.002$t^*$. The simulation is conducted in the canonical ensemble with the temperature controlled by a Langevin thermostat. Equilibrium is reached after $10^6$ simulation steps, and ensemble average quantities are calculated over $10^7$ steps. All MD simulations are performed on the LAMMPS software\cite{plimpton1995fast} (schematically shown in Figure \ref{molecular_system}). It should also noted that we use random-batch Ewald method (RBE)\cite{jin2021random,liang2021superscalability}with slab correction,  in which the empty volume is 3.0 times that of the original system, to speed up the electrostatic interaction calculations\cite{yeh1999ewald}. 



In this study, we evaluate the energy storage capabilities of the supercapacitor by measuring its differential and integral capacitances. The differential capacitance is defined by
\begin{equation} \label{cd}
	C_{dif}=\left(\frac{\partial V}{\partial{\sigma_s}}\right)^{-1},
\end{equation}
\noindent where $\sigma_s$ is the surface charge density, and $V$ is the potential drop from the left electrode $z=-L_z/2$ to the right electrode $z=L_z/2$ ($L_z=30$ in simulations). To obtain the surface charge density, the following modified Poisson's equation is solved for the electrical potential distribution $ \varPhi $:
\begin{equation}
	\varepsilon_c\frac{\mathrm{d}^2 \varPhi}{\mathrm{d} z^2}=-4\pi\Big [ \rho_{net}(z)+\rho_{dipole}(z) \Big ],
	\label{poisson_equ}
\end{equation}
where $\rho_{net}$ and $\rho_{dipole}$ represent the charge density of ions and dipoles, respectively, along the $z$-axis. $\varepsilon_c$ is the dielectric constant.
For a given surface charge $\sigma_s$, the Poisson's equation is solved by a central difference discretization, and the corresponding voltage drop
$V$ is then obtained and the capacitance \eqref{cd} is computed by numerical  differentiation \cite{breitsprecher2014coarse}.
Consequently, the integral capacitance is calculated by \cite{oldham2008gouy}
\begin{equation}
	C_{int}=\frac{1}{V}\int_0^{V}C_{dif}\ \mathrm{d} V.
\end{equation}

\section{Results and discussion}

\begin{figure*}
	\hspace{-8.5em}
	\subfigure{
		\begin{minipage}[t]{0.43\linewidth}
			\centering
			\includegraphics[width=1.45\linewidth]{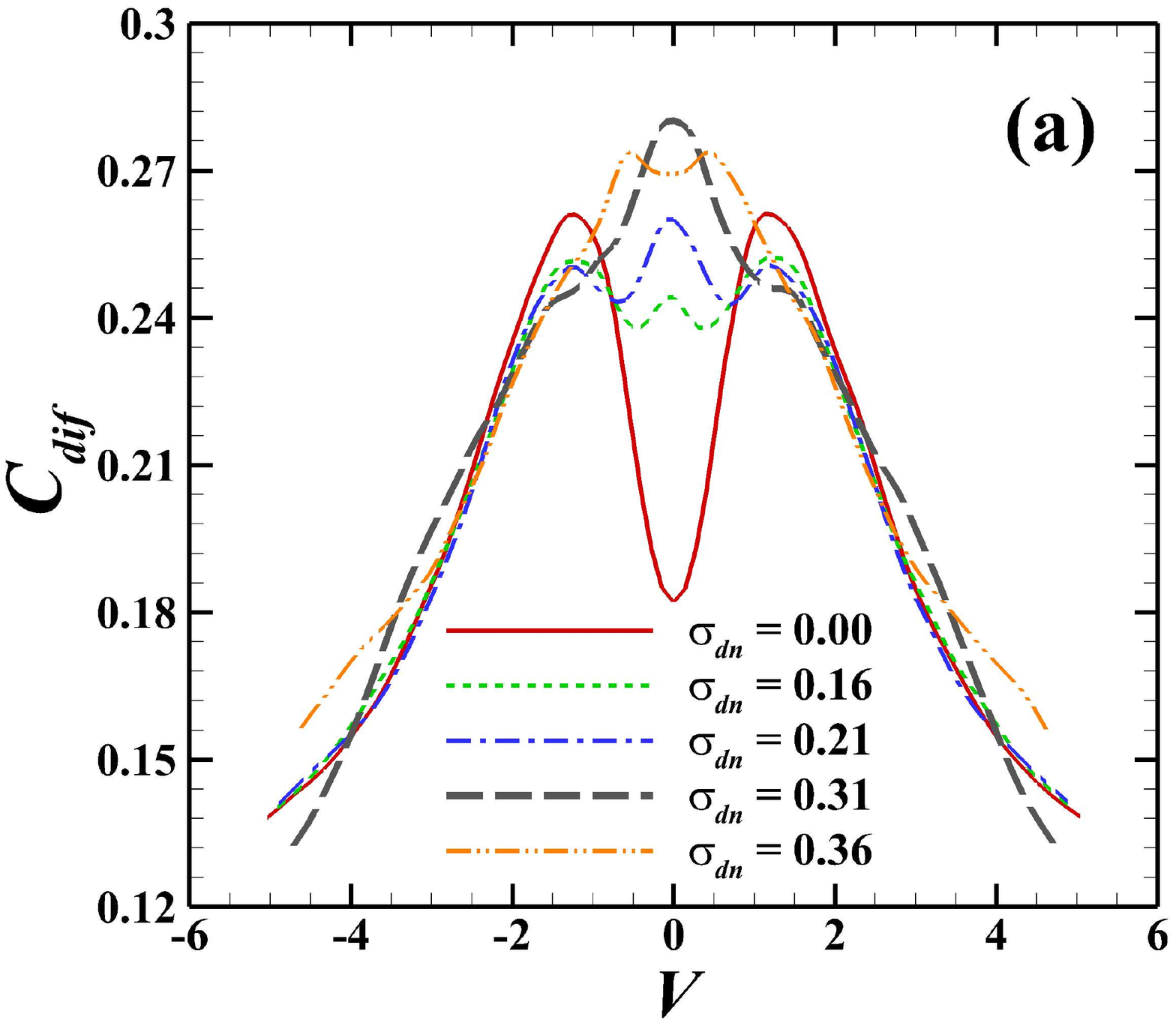} 
			\label{dif_cap}
		\end{minipage}%
	}%
	\hspace{1.8em}
	\subfigure{
		\begin{minipage}[t]{0.44\linewidth}
			\centering
			\includegraphics[width=1.45\linewidth]{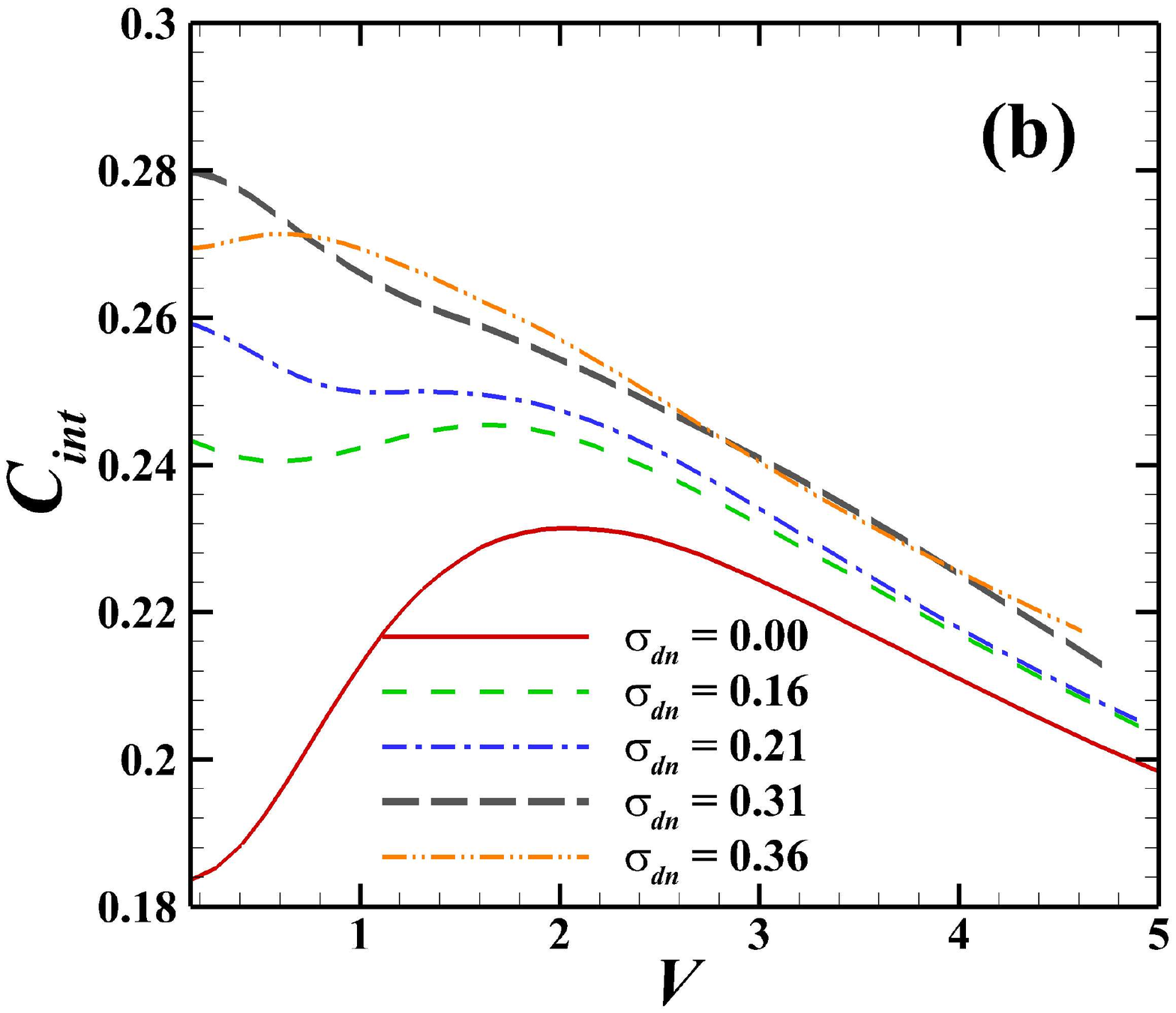}   
			\label{int_cap}
		\end{minipage}%
	}%
	\caption{Capacitance versus potential between two electrodes in three layers of dipoles with respect to different dipole number densities $\sigma_{dn}$. (a) differential capacitance $C_{dif}$, (b) integral capacitance $C_{int}$.}
	\label{cap_dipole}
\end{figure*}

In MD simulations, three layers of dipoles are placed parallel to the surface in the initial state, in which the same number of dipoles are located on each layer and the charges of dipoles are alternatingly arranged. Figure \ref{dif_cap} shows the differential capacitance curves under different surface dipole number densities for an ion volume ratio $v=0.0436$ (corresponding to the salt concentration of 2.918 M). It is well known that when no dipoles coated near the electrode surface, i.e., the dipole number density of each layer ($\sigma_{dn}= 0$), the differential capacitance is camel-shaped. As the ion concentration increases, the capacitance curve
gradually changes from camel shape to bell shape\cite{girotto2018lattice,cats2021differential}. The behavior is explained by the theory of the lattice-gas model as described before. When the dipole number density is gradually increased from 0.16 to 0.31, it is observed that the capacitance shape shows a shape of triple peaks, which means that besides the peak at $V = 0$, there exists two other symmetrical peaks. The curvature between two peaks is getting smoother as $\sigma_{dn}=0.16 \rightarrow 0.31$. When the dipole density increases further, i.e. $\sigma_{dn}= 0.36$, the peak at $V=0$ of the triple peak capacitance returns to the camel shape. 

The corresponding integral capacitance is displayed in  Figure \ref{int_cap}. It is evident that the value of the capacitance with additional dipole layers ($\sigma_{dn}=0.16,0.21,0.31$) is much higher than that of the pure-electrolyte system ($\sigma_{dn}=0$). The capacitance value increases with $\sigma_{dn}$ and reaches the maximum at $\sigma_{dn}= 0.31$. Specifically, at $V=0.2$, the capacitance shows an approximate 55$\%$ increase when comparing the case of $\sigma_{dn}=0.31$ with $\sigma_{dn}=0$. The difference in capacitance between the molecular system containing dipoles and the pure electrolyte simulation system decreases within the range of $V=0.2$ to 2.0. Beyond $V=2.0$, this difference remains unchanged, indicating that the integral capacitance of the dipole-containing system retains an enhancement of almost 10$\%$ compared to the pure electrolyte system. The capacitance starts lowering down when the dipole number density $\sigma_{dn}$ goes beyond a critical value at $\sigma_{dn}=0.31$.
One observes that $C_{int}$ is lower at $\sigma_{dn}=0.36$ than at $\sigma_{dn}=0.31$ for the voltage range $V=0.2\sim0.8$. Then the capacitances for $\sigma_{dn}=0.36$ and $\sigma_{dn}=0.31$ maintain almost the same at $V>0.8$.
By comparing the results in Figure \ref{dif_cap} with Figure \ref{int_cap}, it is indicated that the existence of the triple peak differential capacitance can significantly increase the integral capacitance of the whole system.

Recently, Dong {\it et al.}\cite{dong2023dielectric} developed a dielectric-electrolyte supercapacitor with a stable operating potential up to 3 V for aqueous electrolytes. This progress was achieved through the application of a dielectric polymer layer on the electrode, utilizing the dipole effect. The hydrophobic nature and spatial occupancy effect of the polymer material restrict the interaction between the electrode and water, preventing water decomposition under high electric fields and extending the operating potential range.  Moreover, the polarization of the polymer generates additional active sites for the adsorption of counterions, resulting in an overall enhancement of capacitance.

Modeling the dielectric polymer, specifically polyvinylidene fluoride-co-hexafluoropropylene (PVDF-HFP), presents challenges due to its complex disordered phases \cite{tian2008poly,huan2016advanced}. Additionally, modeling of polymeric materials comprising millions of atoms requires substantial computational time. In this research, we incorporate the spatial effect and the dipolar effect of dipoles into the traditional supercapacitor model. The size effect of dipoles precisely corresponds to the spatial effect of the polymer, while the reorientation of dipoles corresponds to the polarization effect under an electric field. Therefore, the dipole model encompasses these two effects, making it suitable for modeling the polymer. The enhanced integral capacitance profile presented in Figure \ref{int_cap} provides a microscopic theoretical understanding of the experiments. If the electrode materials used for interface modification in supercapacitors meet the conditions  of steric effect and polarization reorientation effects, the dipole model can be employed to investigate the energy storage mechanism of interface modified electrode materials.

\begin{figure*}
%
	 \includegraphics[width=1.0\linewidth]{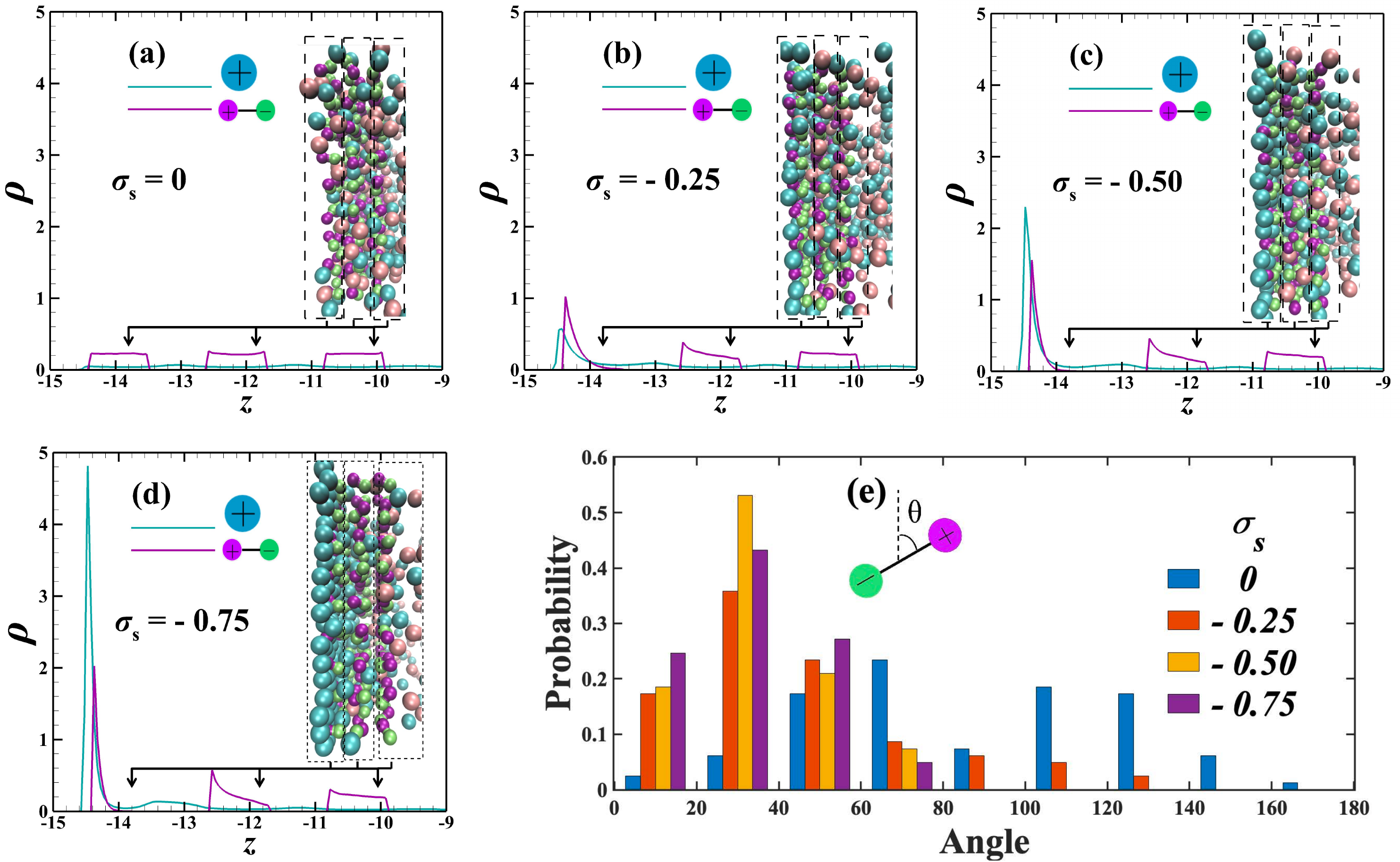}
	\centering
	\caption{Number density profiles of the cation(blue) and the positive segment of dipole(purple) at $\sigma_{dn}=0.21$ under different  surface charge density. (a) $\sigma_s=0 $, (b) $\sigma_s=-0.25 $, (c) $\sigma_s=-0.50 $, (d) $\sigma_s=-0.75 $. A snapshot of particles near the interface is also displayed in each figure. It shows the accumulation of counterions and reorientation of dipoles. (e)     Polarization probability of dipoles versus angle $\theta$ (degree) along $z$ axis at $\sigma_{dn}=0.21$ under different  surface charge density $\sigma_s$.}
	\label{num_density}
\end{figure*}

To gain further insight into the mechanism behind the triple peak capacitance and the improved integral capacitance, we inquire into the EDL structures ($z=-15\sim -9$) for different $\sigma_s$. The number density profile for cations and the positive part of dipoles near the negative electrode are shown in Figure \ref{num_density}. At zero surface charge density ($\sigma_s=0$) corresponding to $V=0$, cations and anions are distributed at the same proportion in Figure \ref{num_density}(a) and dipoles are randomly distributed (Figure \ref{num_density}(e)). The first dipole layer as a whole remains neutral and is called the neutral layer. Meanwhile, the dipoles are not orientated, and thus the added layers of dipoles can be roughly regarded as the new `electrode'. According to the theory of the plane-parallel capacitor, the capacitance is inversely proportional to the distance between electrodes. As a consequence, the capacitance exhibits a peak at $V=0$. As the value of surface charge density $\sigma_s$ increases (i.e. $\sigma_s=-0.25$),  the deviation angle of the first layer ($z=-14.5\sim -13.5 $) along $z$ axis changes significantly (Figure \ref{num_density}(e)), implying that the first dipole layer is reoriented along the direction of potential $V$. 
As a result, the dipoles attract counterions during the reorientation and the positive parts of the dipoles accumulate in the Stern layer ($z=-14.5\sim -14.0 $) (see Figure \ref{num_density}(b)), while coions and negative portions of dipoles are excluded on the other side. This newly formed structure in the first layer is called the bilayer\cite{smith2013monolayer,ivanivstvsev2014poly}. Therefore, the `electrode' made of dipoles decomposes and the capacitance value decreases. Analysis of the case for $\sigma_s$ ranging from 0 to -0.25 reveals a neutral to bilayer transition in the first dipole layer, which is responsible for the triple peak capacitance observed at $V=0$. When $\sigma_s=-0.50$, more counterions are absorbed in the Stern layer as plotted in Figure \ref{num_density}(c, e). When the value of $\sigma_s$ continues to increase (viz. $\sigma_s=-0.75$), the dipoles are almost perpendicular to the electrode. The counterions are massively adsorbed at the electrode because the cation density exceeds more than twice the dipole density in the Stern layer as depicted in Figure \ref{num_density}(d). Simultaneously, the steric effect strengthens, and the bilayer structure is maintained between $\sigma_s=-0.25$ and $-0.75$. In brief, the analysis for the EDL structure demonstrates that the first peak at $V=0$ is caused by the transition from neutral layer to bilayer. 

\begin{table}  \caption{Average accumulated charge of different zones near the cathode, scaled to the absolute value of the surface charge density $\sigma_s$ of the electrode $Q/|\sigma_s|$. The dipole number density $\sigma_{dn}=0.21,0.36$.}
	\vspace{1.0em}
	\begin{tabular}{lcccccc}
		
		\toprule[1.5pt]
		\vspace{0.3em}
		$\sigma_{dn}$ & $\sigma_s$ & zone 1 & zone 2 & zone 3 & zone 4 & zone 5 \\
		
		
	   \midrule[1pt]
		
		& -0.05 & 1.5750 & -0.9326 &0.2151&0.2262 &0.0488    \\
		
		& -0.25 &  1.2729 &-0.5580 &0.1228 &0.1773 &0.0488   \\
		
		0.21 & -0.50 &1.0123 &-0.3306 &0.0813 &0.1267 &0.0380    \\
		
		& -0.75 &0.8045 &-0.2149 &0.0889 &0.1169 &0.0364 \\
		
		 \midrule[1pt]
		
		& -0.05 & 1.7542 & -1.2519 &0.3283 &0.2602 &0.0491    \\
		
		& -0.25 &  1.5480 &-0.9867 &0.2524 &0.2369 &0.0483   \\
		
		0.36 & -0.50 &1.2117 &-0.6082 &0.1625 &0.1835 &0.0456    \\
		
		& -0.75 &0.9565 &-0.4031 &0.1473 &0.1500 &0.0406 \\
		
		\bottomrule[1.5pt]
		
	\end{tabular}
	
	\label{ave_charge_81}
	
\end{table}

To analyse the effect on the capacitance shape caused by the bilayer structure, we calculate the average accumulated charge (scaled by $Q/|\sigma_s|$) near the cathode surface within different zones. Zone 1 corresponds to the area of $z=-14.5\sim-14.0$ and zone 2 corresponds to the area of $z=-14.0\sim-13.5$ and so on.  The upper part of Table \ref{ave_charge_81} shows that the net charge of the Stern layer (viz. zone 1) greatly exceeds  the total surface charge at $\sigma_s=-0.05,-0.25 $ and $-0.50$, which is known as the overscreening phenomenon\cite{fedorov2008towards,fedorov2010double,goodwin2017underscreening}. 
In other words, the transition from neutral layer to bilayer transition gives rise to the overscreening effect. When $\sigma_s=-0.25 \to -0.5$ (corresponding to $V=-1\sim -2 $ in Figure \ref{dif_cap}), counterions are gradually absorbed into the Stern layer and then the overscreening effect is weakened given in Table \ref{ave_charge_81}. Consequently, with the adsorption of counterions near the electrode, the steric effect becomes dominant, giving rise to the second peak.
At $\sigma_s=-0.75$ ($V< -2 $), the overscreening effect disappears and only the steric effect affacts the bilayer strcuture. As a consequence, the differential capacitance shape is the same as in the pure-electrolyte system when $V< -2 $. Therefore, a conclusison can be drawn that the transitions from neutral layer to bilayer and from overscreening effect to steric effect are extremely important for the enhancement of the integral capacitance.

For further analysis, we outline the underlying reasons for the capacitance is no longer triple peak shape when $\sigma_{dn}=0.36$ in Figure \ref{ave_charge_81}. Similar to the analysis in the last paragraph, an overscreening effect arises because the value of the average accumulated charge in zone 1 exceeds 1.0 at $\sigma_s=-0.05,0.25$ and $-0.5$. see the lower part of Table \ref{ave_charge_81}. However, the value of the average accumulated charge in zone 2 at $\sigma_s=-0.05$ exceeds 1.0 likewise, i.e. the second layer of co-ions overcompensates the first layer of counterions. The double overscreening phenomenon results in the steric effect outweighing the overscreening effect at small $\sigma_s$. As a result, the peak arising from the neutral layer to the bilayer at $V=0$ ( when $\sigma_{dn}=0.16,0.21,0.31$) does not appear and thus the capacitance is of camel-shaped.


Experimental findings reported in Dong {\it et al.} \cite{dong2023dielectric} have revealed that the application of voltage across the electrodes induces polarization of the dielectric polymer layer. This polarization leads to the adsorption of ions and subsequent storage of electric energy within the dielectric polymer layer, serving as the primary factor determining the high capacitance value. In our study, we consider the original polymer layer as the neutral dipole layer. The polarization of the dielectric layer gives rise to the formation of active sites, facilitating the development of a bilayer structure. Consequently, this polarization process aligns with the transition from the neutral dipole layer to the bilayer configuration. During the charging process, as the electric field strength increases, the polymer layer shows a considerably stronger attraction towards counterions located near the dipoles. This analysis confirms that the overscreening effect prevails over the steric effect, elucidating the underlying mechanism. Therefore, the observed experimental phenomena provide convincing evidence that both transitions contribute to the notable increase in the integral capacitance.

%

\begin{figure}[htbp]
	\centering
	\includegraphics[width=1.0\linewidth]{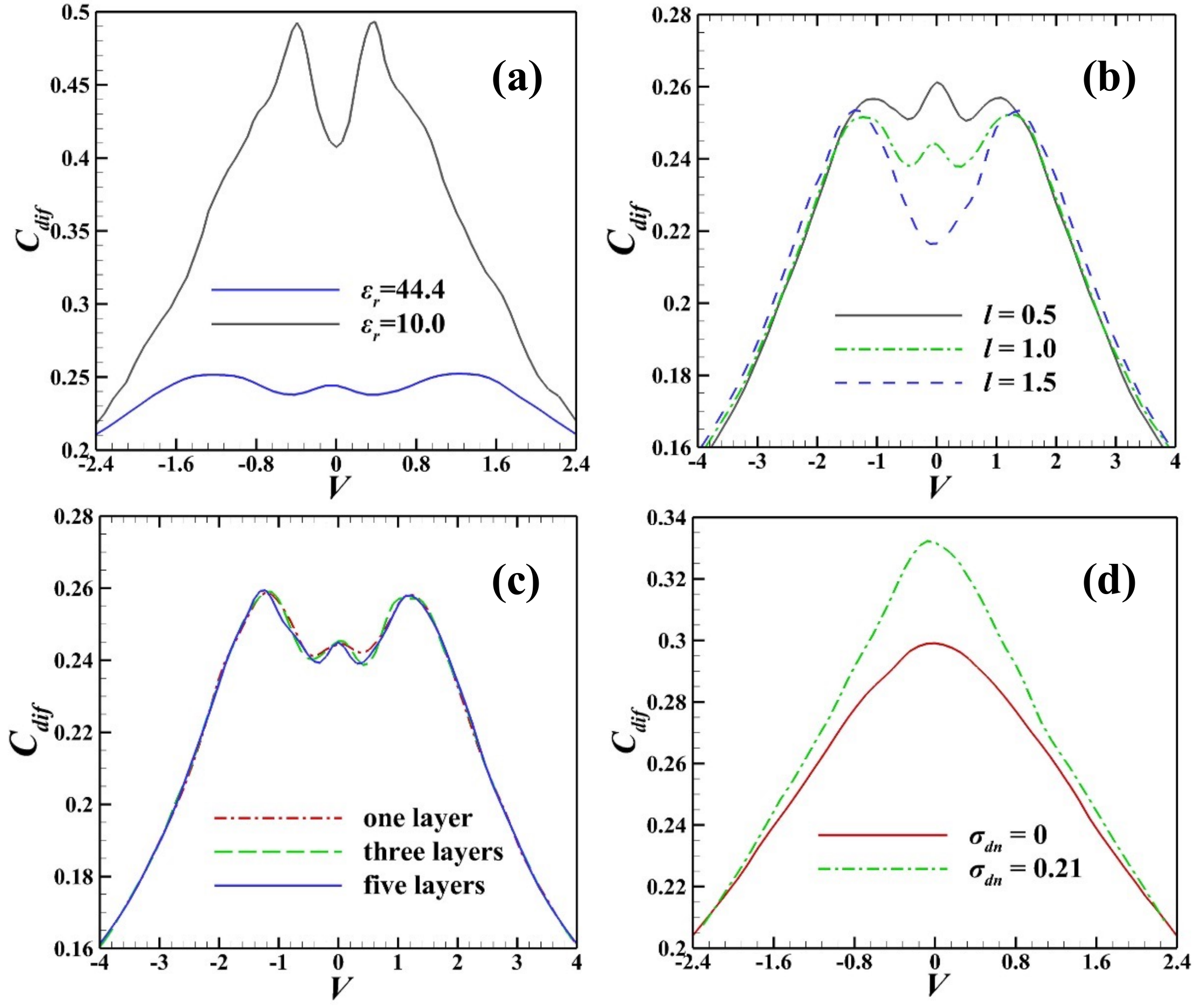}
	\caption{Differential capacitance for different parameters of dipoles. (a) Differential capacitance under different dielectric permittivity, (b) under different intervals $l$ between dipoles, (c) under different layers of dipoles, and (d) under the ion volume ratio $y=0.1308$(8.752M).} 
	\label{parameters}
\end{figure}

In addition, we investigate the behavior of capacitance by varying the parameters of dipoles. In the case of the organic solvent, the differential capacitance exhibits a curve resembling the camel shape (Figure \ref{parameters}(a)). Similarly, the corresponding integral capacitance shows a significant increase compared to the aqueous solution, which is consistent with experimental observations where the capacitance values of organic solvents are higher than those of aqueous solutions. When the intervals between dipoles are varied, the capacitance curve deviates from the triple-peak shape observed at $l=1.5$, instead following a camel-shaped curve (Figure \ref{parameters}(b)). The triple-peak shape is observed at $l=0.5$ and 1.0. The decrease in dipole intervals reduces the available space for counterions near the electrodes, highlighting the importance of dipole reorientation in determining the shape of the differential capacitance curve at lower intervals. As a result, a triple-peak shape is formed as the interval decreases. Comparing the differential capacitance profiles of three layers and five layers as shown in Figure \ref{parameters}(c), we find that the first dipole layer adjacent to the electrode has a dominant influence on the structure of the electric double layer (EDL) and subsequently, the shape of the differential capacitance curve. At higher ion concentrations, the capacitance curve transitions to a bell shape (Figure \ref{parameters}(d)). This transition is attributed to the predominance of spatial effects over the transition from a neutral layer to a bilayer at lower $\sigma_s$, deviating from the triple-peak shape observed in the differential capacitance curve at lower concentrations.

%
%


\section{Conclusion}

In this work, we examine the dipole-induced capacitive effect on the capacitance of supercapacitors by performing MD simulations.  We have found that the differential capacitance shape exhibits triple peaks, while the integral capacitance is over 50$\%$ higher than that of the pure electrolyte system at low voltages, which agrees well with the material based experimental findings. Our findings demonstrate that the reorientation of dipoles plays a key role in the transition from a neutral layer to a bilayer and from an overscreening effect to a steric effect of the counterions, leading to an increase in the integral capacitance. We state that the microscopic mechanism of increasing capacitance is complementary to the energy storage mechanism explained by experiments. Furthermore, we carry out a detailed investigation of the influence of the dipole effect on the capacitance in different systems.
Overall, this work not only clarifies the molecular mechanism of the enhanced capacitance for polymer-coated supercapacitor, but also offers reliable theoretical guidance for the rational design and manufacturing of high performance energy storage devices.

\begin{acknowledgement}

This work is supported by the National Natural Science Foundation of China (grant Nos. 12071288 and 51877132) and the Science and Technology Commission of Shanghai Municipality (grant Nos. 20JC1414100 and 21JC1403700). T. Zhao acknowledges the support from China Postdoctoral Science Foundation (No. 2022M712055). The authors also acknowledge the support from the HPC center of Shanghai Jiao Tong University.

\end{acknowledgement}

%
%
%

\normalem
\bibliography{bib_microscopic.bib}

\end{document}